# Comment on "Electromagnetic Global Gyrokinetic Simulation of Shear Alfvén Wave Dynamics in Tokamak Plasmas" [Phys. Plasmas 14, 042503 (2007)]

Bruce I. Cohen, Andris M. Dimits, and William M. Nevins

University of California, Lawrence Livermore National Laboratory, Livermore,

California 94550

Abstract

This comment clarifies the relation of the research in a recently published article [Phys. Plasmas 14, 042503 (2007)] to other prior publications addressing the inclusion of electromagnetic and drift-kinetic electron physics in gyrokinetic simulation, raises a concern related to the inclusion of kinetic electrons in a system with magnetic shear, and discusses alternatives in the face of an important limitation on the general applicability of the algorithm described therein.

The research published in Ref. 1 addresses extensions of a global, particle-in-cell (PIC), gyrokinetic, simulation algorithm to include the coupling of electrostatic drift-type modes to electromagnetic Alfvénic modes using a hybrid, fluid-kinetic, electron model. This comment seeks to clarify the relation of the research in Ref. 1 to other publications predating Ref. 1, raises a concern related to the inclusion of kinetic electrons [Eq.(12) in Ref. 1] in the presence of mode rational surfaces, and points out some alternatives in the face of an important limitation on the general applicability of the algorithm in Ref. 1.

Reference 1 presents simulations in global geometry with fluid electrons and formulates the perturbative inclusion of kinetic effects for $|\omega/k_\parallel v_e|\ll 1$, but does *not* include electron kinetic effects in any of the simulations reported. Furthermore, in the simulations considered, the parallel vector potential $A_\parallel \neq 0$ and the perpendicular vector potential $A_\perp = 0$. With these simplifications, the underlying physics content of the simulations presented was contained in the hybrid algorithm (fluid electrons and gyrokinetic ions) previously introduced by Parker and Chen[2,3] as noted and cited in Refs. 24 and 25 of Ref. 1.

With respect to the perturbative inclusion of the kinetic electrons in the hybrid algorithm, the work of Cohen, Dimits, Nevins, Chen and Parker[4,5] elaborated and extended a hybrid algorithm that in the absence of magnetic shear is applicable to a class of problems for which $|\omega/k_\parallel v_e| \ll 1$, which is similar to applications that the algorithm in Ref. 1 and Ref. 12 (in Ref. 1) can address. The kinetic electron and electromagnetic algorithm in Ref. 1 was based on Refs. 12 (Lin and Chen) and 13 (Wang, Chen, and Lin) of Ref. 1. The algorithm of Cohen, *et al.*,[4,5] possesses some significant differences, *e.g.*, electron inertia effects are included perturbatively, valid for $|\omega/k_\parallel v_e| \ll 1$, and $A_\parallel$ does not necessarily vanish as $\hat{\mathbf{b}} \cdot \nabla \rightarrow 0$. In contrast, $\partial A_\parallel / \partial t = c\hat{\mathbf{b}} \cdot \nabla(...)$ in Eq.(6) of Ref. 1 and Eq.(9) of Lin and Chen. The approximation required for the treatments in Ref. 1, Lin and Chen, and Cohen, *et al.*, is increasingly valid for increasing values of $\beta m_i / m_e$ ($\beta m_i / m_e > 1$) and invalid for $\beta m_i / m_e < 1$. It is of interest to note that the simulations in systems with no magnetic shear, addressing shear-Alfvén waves in Lin and Chen (Fig. 1) and in Cohen, *et al.*, and collisionless drift and ion-temperature gradient instabilities in Cohen, *et al.*, yield relatively accurate results when compared to linear theory for $\beta m_i / m_e \geq 1$ and become inaccurate for $\beta m_i / m_e < 1$.

The algorithm introduced in Refs. 4 and 5 has been demonstrated successfully only in an unsheared magnetic field (no mode rational surfaces). In contrast, the GEM code that uses an electromagnetic, particle-in-cell, split-weight, kinetic-electron algorithm has been shown to work in a sheared magnetic field; and the algorithm is not limited to $|\omega/k_\parallel v_e|<<1$ nor is it restricted in the polarity of the modes with respect to the mode rational surface.[6,7] With magnetic shear, algorithms based on the fundamental assumption $|\omega/k_\parallel v_e|<<1$ can become invalid near mode rational surfaces where $k_\parallel \Rightarrow 0$ unless the parallel electric field is so small that even though the model produces an incorrect electron response in the region near the mode rational surface, the response is so small as to be negligible. Otherwise the incorrect electron response produces errors in the electron charge and current densities that then contaminate the Maxwell equations that determine the fields self-consistently. We do not know whether the electromagnetic algorithm in Ref. 1 will necessarily fail when kinetic electron effects are included and mode rational surfaces are present because the algorithm in Ref. 1 forces $A_\parallel$ to vanish at the mode rational surfaces. However, it is reasonable to ask how the electromagnetic algorithm in Ref. 1 deals with the electron resonance layer and the mode rational surfaces when kinetic electrons are present.

Reference 1 recognizes that $|\omega/k_\parallel v_e|<<1$ is an essential constraint for the algorithm presented there and limits the applicability of this extended hybrid algorithm to modes such that $A_\parallel$ has odd parity and $\phi$ has even parity with respect to the mode rational surfaces. The parity restriction is built into the algorithm in Ref. 1. The parity restriction also implies a limitation on the parity of the nonlinearities that the model in Ref. 1 can support. These constraints and limitations evidently cause no difficulties in simulations with only fluid electrons, but no indication is given in Ref. 1 nor in Lin and Chen (the Wang, Chen, and Lin citation is an abstract) of how the electromagnetic algorithm actually performs when there are mode rational surfaces and kinetic electrons. In the algorithm proposed in Ref. 1, $A_\parallel$ goes to zero at the mode rational surface but can be non-zero nearby. Given that there is a finite region near the mode rational surface in which the electron response is incorrectly modeled because $|\omega/k_\parallel v_e|$ can be large therein, the resulting perturbed electron number density and perturbed current density will be incorrect in this domain. However, does this matter? In this circumstance, the algorithm can still yield correct results if the electron number and current densities are negligibly small in this domain and have no significant influence on the self-consistent fields. Whether this is realized and the algorithm remains valid is not addressed adequately in Ref. 1 nor in Lin-Chen. A simulation demonstration with mode rational surface(s) and kinetic electrons would be enlightening. More generally, it would be desirable to compare simulation results using this electromagnetic algorithm with kinetic electrons when mode rational surfaces are present to simulations using a model *not* restricted to the constraint $|\omega/k_\parallel v_e|<<1$ and not subject to the parity restriction. If such comparisons were routinely required when undertaking large global microturbulence simulations to establish confidence in the results of the algorithm described in Ref. 1 when kinetic electrons were included, this could be burdensome for large, compute-intensive simulation series.

      The limitation on $|\omega/k_\parallel v_e|<<1$ and the restriction on the parity of the electromagnetic fields in Ref. 1 are not required in other electromagnetic, kinetic-electron algorithms, which have been implemented in toroidal geometry for general equilibria, and

have proven to be robust.[6-10] Besides the particle-in-cell algorithm in Refs. 6 and 7, there are toroidal, electromagnetic, gyrokinetic, continuum algorithms with kinetic electrons. There have been successful code cross-checks between the particle code in Refs. 6 and 7, and the continuum codes listed in 8 and 10. References 4, 5, 6, 7, 8, 9, and 10 are not cited in Ref. 1. Moreover, toroidal electromagnetic simulations[11,12] undertaken with the GYRO code[10] using a more general kinetic electron model than in Ref. 1 showed structure in the electron response centered at the mode rational surfaces, *e.g.*, localized electron temperature corrugations that have been interpreted as arising from the electron resonance layers near the mode rational surfaces. Refs. 11 and12 lend support to the concern raised in this Comment regarding the significance of the resonant electron response near the mode rational surface. A variant of the GEM algorithm employing drift-kinetic electrons and fluid ions was used to perform two-dimensional electromagnetic gyrokinetic simulations of the collisionless and semi-collisional tearing modes, and comparisons were made to a linear eigenmode analysis.[13] The GEM particle simulations resolved the electron resonance layer near the mode rational surface to obtain correct results. Reference 13 demonstrates that the GEM electromagnetic algorithm (one of the more general alternatives to the algorithm in Ref. 1) can resolve tearing modes and the electron resonance layer.

Reference 1 is not a review article and introduces its own specific equations capturing the toroidal extension of the Lin and Chen algorithm. Part of the motivation for this Comment is to place the work of Ref. 1 in the larger context of other publications (Refs. 4, 5, 6, 7, 8, 9, and 10) that also address the subject of the inclusion of drift-kinetic electrons and electromagnetic effects in gyrokinetic simulation, to better inform the readership of the journal.

This work performed under the auspices of the U.S. Department of Energy by the Lawrence Livermore National Laboratory under Contract DE-AC52-07NA27344.   We thank Scott Parker and Yang Chen for allowing us to use their GEM code and Jeff Candy, Greg Hammett, and Ron Waltz for illuminating discussions.